# Deep Learning for Longitudinal Gross Tumor Volume Segmentation in MRI-Guided Adaptive Radiotherapy for Head and Neck Cancer

Xin Tie [0000-0003-3062-5995], Weijie Chen [0000-0002-1608-0899], Zachary Huemann [0000-0002-1472-243X], Brayden Schott [0009-0000-3326-1854], Nuohao Liu [0000-0003-3526-0980] and Tyler J. Bradshaw [0000-0001-9549-7002]

University of Wisconsin, Madison, WI, USA

**Abstract.** Accurate segmentation of gross tumor volume (GTV) is essential for effective MRI-guided adaptive radiotherapy (MRgART) in head and neck cancer. However, manual segmentation of the GTV over the course of therapy is time-consuming and prone to interobserver variability. Deep learning (DL) has the potential to overcome these challenges by automatically delineating GTVs. In this study, our team, *UW LAIR*, tackled the challenges of both pre-radiotherapy (pre-RT) (Task 1) and mid-radiotherapy (mid-RT) (Task 2) tumor volume segmentation. To this end, we developed a series of DL models for longitudinal GTV segmentation. The backbone of our models for both tasks was SegResNet with deep supervision. For Task 1, we trained the model using a combined dataset of pre-RT and mid-RT MRI data, which resulted in the improved aggregated Dice similarity coefficient (DSC$_{agg}$) on a hold-out internal testing set compared to models trained solely on pre-RT MRI data. In Task 2, we introduced mask-aware attention modules, enabling pre-RT GTV masks to influence intermediate features learned from mid-RT data. This attention-based approach yielded slight improvements over the baseline method, which concatenated mid-RT MRI with pre-RT GTV masks as input. In the final testing phase, the ensemble of 10 pre-RT segmentation models achieved an average DSC$_{agg}$ of 0.794, with 0.745 for primary GTV (GTVp) and 0.844 for metastatic lymph nodes (GTVn) in Task 1. For Task 2, the ensemble of 10 mid-RT segmentation models attained an average DSC$_{agg}$ of 0.733, with 0.607 for GTVp and 0.859 for GTVn, leading us to achieve 1st place. In summary, we presented a collection of DL models that could facilitate GTV segmentation in MRgART, offering the potential to streamline radiation oncology workflows.

## 1      Introduction

Radiation therapy (RT) is a cornerstone of cancer treatment, particularly for head and neck cancer (HNC). Over the past decades, the treatment of HNC with RT has seen significant advancements, evolving from 3D conformal radiation therapy (CRT) into intensity-modulated radiation therapy (IMRT) [1]. IMRT enables improved targeting of tumors while sparing normal tissues. However, this conformality also poses a critical challenge: anatomical changes during treatment, such as tumor shrinkage or weight loss, can drastically alter the dose delivered to both tumor and surrounding organs-at-risk. To address this, adaptive radiation therapy (ART), which involves re-planning during treatment in response to changes taking place in the patient's body, has been developed with the goal of improving target coverage and reducing normal tissue toxicity. Recently, MRI-guided adaptive radiotherapy (MRgART) has emerged as a promising approach for treating HNC patients due to the superior soft tissue contrast provided by MRI, allowing for more accurate tumor delineation [2]. However, manual segmentation of gross tumor volume (GTV) on pre- and mid-treatment MRI scans is usually time-consuming and subject to inter-observer variability. Artificial intelligence (AI), especially deep learning (DL), has the potential to streamline this process, facilitating timely and accurate adjustments to treatment plans.

Extensive studies have focused on segmenting tumors across various imaging modalities using DL [3], [4], [5], [6], [7]. Despite these advances, there remains a gap in the development of DL tools for segmenting tumors on multi-time-point imaging data, which is crucial for ART. The annual Medical Image Computing and Computer Assisted Intervention Society (MICCAI) Head and Neck Tumor Segmentation for MR-Guided Applications (HNTS-MRG) 2024 challenge addressed this gap by releasing high-quality annotated MRI data and promoting the development of DL models capable of segmenting GTVs on MRI at different treatment stages. This challenge includes two tasks: Task 1 focuses on automatically segmenting tumor volumes on pre-RT MRI, while Task 2 targets the segmentation of tumor volumes on mid-RT MRI.

In this work, we introduced and validated a series of DL methods designed for longitudinal GTV segmentation in MRgART. Notably, model predictions on current timepoint scans are guided by previous timepoint scan information (when applicable) via specialized attention mechanisms. For each task, we reported the results for ablations studies to understand the impact of each carefully selected component.

## 2      Datasets and Methods

### 2.1      Imaging Datasets

The retrospective training dataset consists of 150 patients, each having pre- and mid-RT T2-weighted MRI scans and corresponding labels for primary GTV (GTVp) and metastatic lymph nodes (GTVn). All the cases were manually segmented by multiple physicians independently. The final ground truth contours were combined via the STAPLE algorithm [8] and verified by experienced radiation oncologists. For Task 2, all the pre-RT data has been deformably registered to mid-RT MRI for better spatial alignment. The training data was provided in Neuroimaging Informatics Technology Initiative (NIfTI) format.

## 2.2    Model Architecture

For both segmentation tasks, we employed SegResNet [9] with deep supervision (**Fig. 1**) as the backbone of our models. This architecture has demonstrated consistently high performance in previous challenges [10], making it a reliable choice for our study.

SegResNet is a convolutional encoder-decoder model initially designed for brain tumor segmentation on MRI. The encoder is composed of multiple stages, each containing several convolutional blocks with residual connections [11]. Our model's architecture starts with a single residual block in Level 1, followed by progressively deeper configurations of 2, 2, 4, 4, and 4 residual blocks in the subsequent levels. Each residual block is a stack of two units, where each unit includes instance normalization, ReLU activation, and a 3×3×3 convolution. To effectively capture multi-scale contextual information, we downsample the feature maps by a factor of 2 at each level, while simultaneously increasing the number of feature channels.

The decoder mirrors the encoder's structure, and each level contains a single convolutional block. To reconstruct the pixel-wise segmentation masks, we upsample the feature maps using transposed convolutions and reduced the number of feature channels between levels. Before passing these features into the decoder's convolutional block at each level, we fuse them with the output features from the encoder at the same spatial level. The last layer is a 1×1×1 convolution, which reduces the channels to three output channels, followed by a softmax function to estimate the probability of each pixel belonging to each class (i.e., background, GTVp and GTVn).

**Fig. 1.** Diagram of the SegResNet architecture with deep supervision.

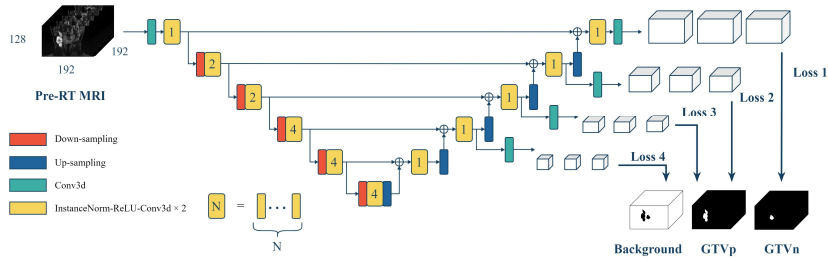

For Task 1, the model input is an MRI image from a single time point. For Task 2, the model takes a mid-RT MRI image along with corresponding pre-RT GTVp and GTVn masks as inputs. To allow prior information from pre-RT data to influence intermediate features in the mid-RT model, we integrated a mask-aware attention module at each level of the encoder (**Fig. 2**). This module is an adaptation of the convolutional block attention module (CBAM) [12].

**Fig. 2.** (**A**) shows the SegResNet architecture augmented with mask-aware attention modules. (**B**) illustrates the inner workings of mask-aware attention.

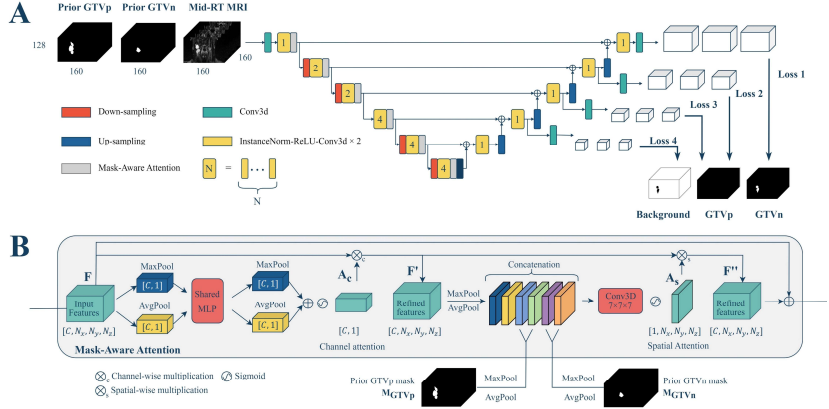

Similar to CBAM, the mask-aware attention module first applies channel attention by performing global averaging pooling and global max pooling across the spatial dimensions, feeding pooled features into a shared multi-layer perceptron and generating channel attention weights using a sigmoid activation function. The input feature map is multiplied by these attention weights to emphasize important channels. The channel attention sub-module can be summarized mathematically as follows:

$$A_c = \sigma\left(MLP\left(AvgPool_s(F)\right) + MLP\left(MaxPool_s(F)\right)\right) \quad (1)$$

$$F' = A_c \otimes_c F \quad (2)$$

Where $F$ denotes the input feature, $AvgPool_s$ and $MaxPool_s$ are the pooling operations along the spatial dimensions, $MLP$ is the multi-layer perceptron, $\sigma$ denotes the sigmoid function, $A_c$ denotes the channel attention weights, $\otimes_c$ represents channel-wise multiplication, and $F'$ represents the refined feature.

Next, the mask-aware attention applies spatial attention by performing averaging pooling and max pooling along the channel axis, then concatenating the resulting feature maps with the masks derived from pre-RT GTVp and GTVn masks. To align the spatial dimensions of the features and pre-RT masks at intermediate layers, we applied max pooling and average pooling to both GTVp and GTVn masks progressively along each spatial axis with a kernel size of 3×3×3 and a stride of 2. The spatial attention sub-module involves the following operations:

$$A_s = \sigma\left(f_{7 \times 7 \times 7}\begin{bmatrix} AvgPool_c(F'), & MaxPool_c(F'), \\ AvgPool_s^{(i)}\left(M_{GTV_p}\right), & MaxPool_s^{(i)}\left(M_{GTV_p}\right), \\ AvgPool_s^{(i)}\left(M_{GTV_n}\right), & MaxPool_s^{(i)}\left(M_{GTV_n}\right) \end{bmatrix}\right) \quad (3)$$

$$F'' = A_s \otimes_s F' \quad (4)$$

Where $F'$ represents the channel attention-refined feature, $AvgPool_c$ and $MaxPool_c$ are the pooling operations along the channel axis, $M_{GTV_p}$ and $M_{GT_n}$ are the binary masks of GTVp and GTVn in pre-RT MRI images, $AvgPool_s^{(i)}$ and $MaxPool_s^{(i)}$ indicate $i$ iterations of pooling operations to match the output GTV masks with the feature maps in the spatial dimensions, $[\cdot]$ indicates feature concatenation, $f_{7\times7\times7}$ denotes a $7\times7\times7$ convolution with the number of filters equal to 6, $A_s$ denotes the spatial attention weights, $\otimes_s$ represents spatial-wise multiplication. The final refined feature, $F''$, are then added to the input feature $F$:

$$F_{out} = F + F'' \tag{5}$$

This modification at the intermediate layers of SegResNet allows the mid-RT segmentation model to better focus on important regions informed by the pre-RT data. To distinguish it from the original SegResNet, we refer to this architecture as Mask-Aware SegResNet (MA-SegResNet).

### 2.3 Implementation

For both tasks, we resampled the MRI images and corresponding GTV masks to a fixed isotropic voxel size of 1 mm. The MRI intensity was then rescaled using z-score normalization, where only non-zero values were used to compute the mean and standard deviation for each MRI volume. The entire dataset was randomly split into 5 folds, with each fold containing 30 patients (60 scans total, as each patient had both pre-RT and mid-RT imaging data). When we ran fivefold cross-validation, data from four folds was combined for model training, while the remaining fold was used for validation.

In Task 1, the SegResNet model operates on 192×192×128 patches cropped from MRI images centered on the foreground classes with probabilities of 0.45 for GTVp and 0.45 for GTVn (0.1 for background). Both pre-RT and mid-RT MRI data were used for model training, but only pre-RT data in the validation set was used for model selection.

In Task 2, the MA-SegResNet model processes 160×160×128 patches centered on the foreground classes with the same probabilities as in Task 1. For both tasks, the optimal in-plane patch size was selected based on tuning across 192×192, 160×160 and 128×128. In addition to mid-RT MRI images, registered pre-RT GTV masks were included as input. The original GTV masks for pre-RT MRI images, deformably registered to mid-RT MRI images, were first converted to one-hot encoded masks, and only the masks corresponding to GTVp and GTVn were retrieved and used as additional input channels. When training the mid-RT segmentation model, we also included paired pre-RT data as additional training samples. Since no prior information is available for pre-RT MRI images, we set the input GTVp and GTVn masks to zeros. For model selection, only mid-RT data pairs in the validation set were used.

To alleviate the model overfitting problem, we applied the same data augmentation strategies across both tasks, including random affine transformation (rotation between -25 and 25 degrees, axis flip for all dimensions, zoom between 0.8 to 1.2), Gaussian noise, and Gaussian blur. The loss function for both tasks is a compound loss, comprised of cross-entropy and Dice loss:

$$L = \sum_{k=1}^{4} \frac{1}{2^{k-1}} \sum_{j=1}^{N} [L_{CE}(y_j^{(k)}, \hat{y}_j^{(k)}) + L_{DICE}(y_j^{(k)}, \hat{y}_j^{(k)})] \qquad (6)$$

Where $\hat{y}_j^{(k)}$, $y_j^{(k)}$ denote the prediction and the ground truth for the j-th sample at the deep supervision level k, $L_{CE}$ represents the cross-entropy loss and $L_{DICE}$ represents the Dice loss. When $k = 1$, the loss is computed at the same spatial level as the input images. For $k > 1$, the loss is computed at a spatial level with dimensions reduced to $\frac{1}{2^{k-1}}$ of the input image dimension, and the ground truth masks are interpolated using nearest-neighbor interpolation to match the reduced dimensions.

Both pre-RT and mid-RT segmentation models were trained using the AdamW optimizer [13], with an initial learning rate of $10^{-4}$, weight decay regularization of $10^{-5}$, and a cosine annealing scheduler. We randomly sampled 2 patches from each sample and set the batch size to 3 (6 patches in a single batch). The model was trained for 400 epochs on a single NVIDIA A100 GPU. Batch size and number of epochs were optimized via grid search (batch sizes of 2, 3, and 4; epochs of 300, 400, and 500). The learning environment requires the following Python (3.8.8) libraries: PyTorch (1.13.0), MONAI (1.3.0) [14].

For both tasks, we used three different random seeds for model training in each training/validation split. The model with the highest aggregated Dice similarity coefficient (DSC$_{agg}$) in the validation set was selected for each fold, resulting in 5 models for five-fold cross-validation. We repeated the experiments twice with different random seeds to generate the five folds, yielding a total of 10 models. The final submission for each task is an ensemble of these 10 models.

## 2.4    Inference

For each case, we employed the sliding window method with an overlap rate of 0.625 and blended outputs of overlapping patches using Gaussian weighting. Then we averaged the probability maps estimated by 10 individual models. The class label (0: background, 1: GTVp, 2: GTVn) with the highest averaged probability across all three classes was then assigned to each voxel. In Task 1, we removed any small regions with a volume below $0.5$ cm$^3$ using connected component analysis. In Task 2, we implemented a technique that excludes predicted contours on mid-RT MRI scans that have no overlapping voxels with registered pre-RT GTV contours since the mid-RT GTV is expected to shrink and remain confined within the pre-RT GTV contours throughout the course of treatment. This technique, known as "mask propagation through deformable registration" (or MPDR), has been used in previous literature [7] to reduce false positives.

## 2.5    Ablation Studies

We conducted ablation studies to identify key factors that contribute to improved segmentation accuracy. First, we held out 30 cases for testing and altered various settings during model development. The training dataset, comprising 120 cases in total, was divided into five folds; in each iteration, four folds were used for training and one fold for validation. Models trained on these five folds were then evaluated on the hold-out testing set to assess the impact of different configurations on performance.

In Task 1, we investigated whether including paired mid-RT data could enhance the performance of pre-RT segmentation models. In Task 2, we focused on evaluating different combinations of input data, the usefulness of pre-RT data for mid-RT GTV segmentation, the impact of model architectures, and whether applying the MPDR method to the model predictions could further improve performance.

For statistical analysis, we employed the bootstrap resampling technique. In each iteration, we randomly sampled 30 cases from the internal test set with replacement and calculated the $DSC_{agg}$ for each model. The $DSC_{agg}$ values were then averaged across the five models to yield a single metric for each bootstrap trial. This process was repeated 10,000 times. The difference between two model configurations was considered statistically significant at the 0.05 level if the metric value computed for one configuration exceeded that of the other in 95% of trials.

### 2.6 Model Availability

The code and model weights have been made available in an open-source project: https://github.com/xtie97/HNTS-MRG24-UWLAIR.

## 3 Results

### 3.1 Quantitative Results

**Table 1** presents the results of fivefold cross-validation for Task 1, reported as $DSC_{agg}$ for different fivefold splits (split 1 and split 2 were created using different random seeds for data partitioning). The average $DSC_{agg}$ for split 1 is 0.814, and for split 2, it is 0.816. The overall average $DSC_{agg}$ for the 10 models used in the submission is 0.815 across the validation sets. In the final testing phase, the ensemble of these 10 models achieved an average $DSC_{agg}$ of 0.794 on the hold-out 50 cases, with 0.745 for GTVp and 0.844 for GTVn. The drop in $DSC_{agg}$ is primarily attributed to the decreased performance in the GTVp segmentation.

**Table 1.** Cross validation results for pre-RT GTV segmentation (Task 1)

| | **Fivefold Split 1** | | | **Fivefold Split 2** | | |
|---|---|---|---|---|---|---|
| | $DSC_{agg}$ (GTVp) | $DSC_{agg}$ (GTVn) | Average $DSC_{agg}$ | $DSC_{agg}$ (GTVp) | $DSC_{agg}$ (GTVn) | Average $DSC_{agg}$ |
| **Fold 1** | 0.789 | 0.837 | 0.813 | 0.746 | 0.838 | 0.792 |
| **Fold 2** | 0.815 | 0.844 | 0.829 | 0.792 | 0.882 | 0.837 |
| **Fold 3** | 0.812 | 0.881 | 0.846 | 0.783 | 0.853 | 0.818 |
| **Fold 4** | 0.718 | 0.820 | 0.769 | 0.799 | 0.860 | 0.829 |
| **Fold 5** | 0.763 | 0.861 | 0.812 | 0.802 | 0.812 | 0.807 |
| **Average** | 0.779± 0.040 | 0.849± 0.023 | 0.814± 0.029 | 0.784± 0.023 | 0.849± 0.026 | 0.816± 0.018 |

**Table 2** shows the cross-validation results for Task 2 with different fivefold splits. The average DSC$_{agg}$ of the 10 mid-RT models is 0.754 across the validation sets. In the final testing phase, the ensemble of these models attained an average DSC$_{agg}$ of 0.733 on the hold-out mid-RT data, with 0.607 for GTVp and 0.859 for GTVn. Similar to Task 1, the slight decrease in DSC$_{agg}$ is mainly due to the performance drop of the GTVp segmentation.

**Table 2.** Cross validation results for mid-RT GTV segmentation (Task 2)

| | **Fivefold Split 1** | | | **Fivefold Split 2** | | |
|---|---|---|---|---|---|---|
| | DSC$_{agg}$ (GTVp) | DSC$_{agg}$ (GTVn) | Average DSC$_{agg}$ | DSC$_{agg}$ (GTVp) | DSC$_{agg}$ (GTVn) | Average DSC$_{agg}$ |
| **Fold 1** | 0.620 | 0.834 | 0.727 | 0.599 | 0.823 | 0.711 |
| **Fold 2** | 0.714 | 0.834 | 0.774 | 0.679 | 0.868 | 0.774 |
| **Fold 3** | 0.597 | 0.851 | 0.724 | 0.685 | 0.860 | 0.772 |
| **Fold 4** | 0.657 | 0.822 | 0.739 | 0.702 | 0.844 | 0.773 |
| **Fold 5** | 0.728 | 0.877 | 0.802 | 0.651 | 0.841 | 0.746 |
| **Average** | 0.663± 0.057 | 0.843± 0.021 | 0.753± 0.034 | 0.663± 0.040 | 0.847± 0.018 | 0.755± 0.027 |

**Fig. 3** and **Fig. 4** display the example cases from pre-RT MRI and mid-RT MRI scans, respectively. The ground truth and predicted GTV contours are both overlaid on the MRI images. For mid-RT cases, the registered pre-RT scans are also included to reference the initial tumor locations.

**Fig. 3.** Pre-RT segmentation results for six sample cases. Dice similarity coefficients (DSCs) for both GTVp and GTVn are reported above each example.

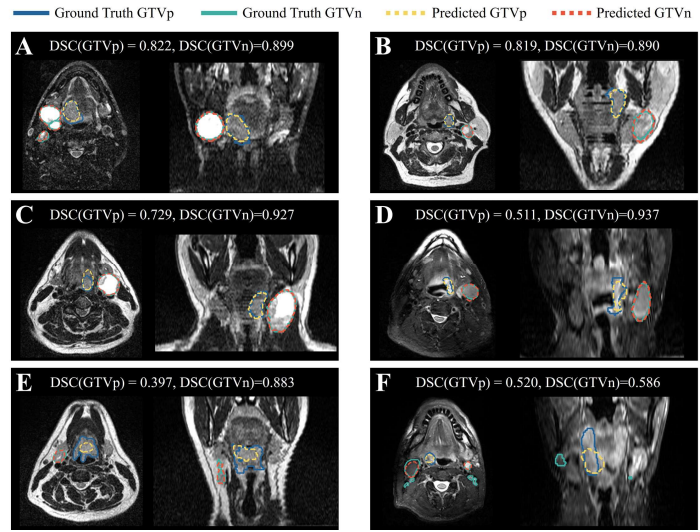

**Fig. 4.** Mid-RT segmentation results for five sample cases. Dice similarity coefficients (DSCs) for both GTVp and GTVn in midRT MRI scans are labeled above each example. Registered pre-RT MRI images are provided to reference initial tumor sites.

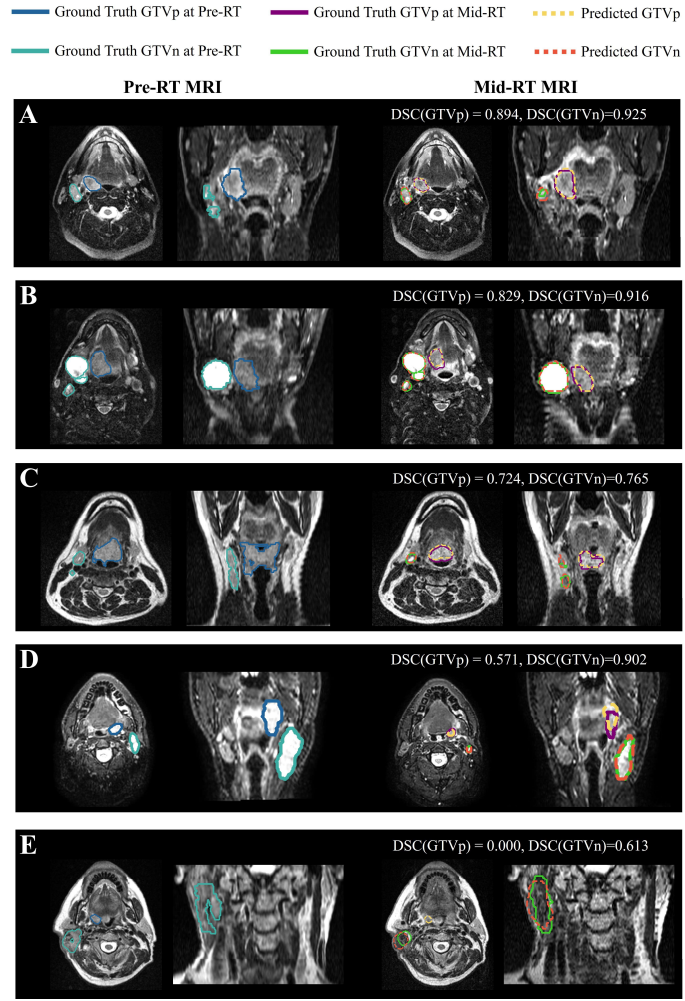

## 3.2 Ablation Studies

The internal testing results in terms of DSC$_{agg}$ for Task 1 are presented in **Table 3**. When paired mid-RT data was included in training, the DSC$_{agg}$ averaged over the five folds was 0.799, which is significantly higher (P=0.004) than the DSC$_{agg}$ achieved by models trained solely on pre-RT data (0.776). This finding supports our decision to train models on data from both time points while only using pre-RT data for model selection.

**Table 3.** Ablation study investigating the impact of training data on the pre-RT GTV segmentation (Task 1)

| Training Data | Fold 1 | Fold 2 | Fold 3 | Fold 4 | Fold 5 | Average |
|---|---|---|---|---|---|---|
| Pre-RT only | 0.770 | 0.789 | 0.778 | 0.774 | 0.770 | 0.776±0.008 |
| Mid-RT & Pre-RT | **0.793** | **0.814** | **0.799** | **0.807** | **0.784** | **0.799±0.012** |

**Table 4** shows the results of ablation studies conducted for Task 2. When no prior information from pre-RT scans was incorporated into mid-RT GTV segmentation, the $DSC_{agg}$ on the internal testing set averaged over the five folds was 0.588. Including paired pre-RT data in model training – without directly utilizing pre-RT information for mid-RT segmentation – resulted in consistent improvements (P=0.001) in $DSC_{agg}$.

**Table 4.** Ablation study investigating the impact of training data, model architectures, post-processing techniques on the mid-RT GTV segmentation (Task 2)

| Input Data | Training Data | Model Architecture | Post-Processing | Fold 1 | Fold 2 | Fold 3 | Fold 4 | Fold 5 | Average |
|---|---|---|---|---|---|---|---|---|---|
| MRI images | Mid-RT only | SegResNet | None | 0.570 | 0.558 | 0.608 | 0.604 | 0.602 | 0.588±0.023 |
| MRI images | Mid-RT & Pre-RT | SegResNet | None | 0.621 | 0.635 | 0.619 | 0.646 | 0.645 | 0.633±0.013 |
| Mid-RT MRI & Pre-RT MRI & Pre-RT GTV masks | Mid-RT only | SegResNet | None | 0.696 | 0.662 | 0.713 | 0.698 | 0.694 | 0.693±0.019 |
| MRI images & Prior GTV masks | Mid-RT only | SegResNet | None | 0.700 | 0.713 | 0.714 | 0.707 | 0.686 | 0.704±0.012 |
| MRI images & Prior GTV masks | Mid-RT & Pre-RT | SegResNet | None | **0.723** | 0.685 | 0.721 | 0.714 | 0.719 | 0.712±0.016 |
| MRI images & Prior GTV masks | Mid-RT & Pre-RT | MA-SegResNet | None | 0.717 | 0.723 | 0.746 | 0.726 | 0.713 | 0.725±0.013 |
| MRI images & Prior GTV masks | Mid-RT & Pre-RT | MA-SegResNet | MPDR | 0.722 | **0.733** | **0.751** | **0.730** | **0.731** | **0.733±0.011** |

The most significant performance increase from the baseline configuration (i.e., using only mid-RT data for model development) occurred when pre-RT GTV masks were integrated. In this setup, simply concatenating mid-RT MRI images with pre-RT GTV masks as model inputs improved the $DSC_{agg}$ from 0.588 to 0.704 (P<0.001). Building on it, further improvements were observed by including pre-RT scans as additional training data (with prior GTV masks set to zeros) and replacing SegResNet model with MA-SegResNet, leading to a $DSC_{agg}$ of 0.725 (from 0.704, P=0.039). Lastly, applying MPDR to the predictions of the MA-SegResNet model consistently increased the $DSC_{agg}$ across all five folds. The configuration in the last row of **Table 4** was the setup used in our final submission for Task 2.

## 4    Discussion

In this work, we addressed both pre-RT and mid-RT MRI segmentation tasks using the SegResNet architecture as the backbone. Our results demonstrated that including paired data from different time points in the treatment as additional training samples enhanced segmentation performance across both tasks. For mid-RT GTV segmentation, integrating prior information from pre-RT scans significantly improved accuracy. Moreover, architectural modifications and post-processing techniques led to further improvements without adding excessive computational complexity and inference time.

There are several limitations in our study. First, we did not incorporate co-registered pre-RT MRI images as an additional input channel for mid-RT GTV segmentation. Adding this information may lead to further improvements. Second, in Task 2, we did not investigate whether the mask-aware attention modules could enhance performance with backbone architectures other than SegResNet. Third, the applicability of our findings and approaches to other longitudinal tumor segmentation tasks in radiotherapy remains uncertain. Lastly, although our method shows clear advancements, its impact on clinical efficiency is unknown. Prospective studies are essential to rigorously evaluate any AI models intended for clinical use.

In conclusion, we developed a series of DL models for automatic GTV contouring in MRgART, demonstrating potential to streamline radiation oncology workflows for treating HNC patients.

**Acknowledgments.** We acknowledge the organizers of the HNTS-MRG 24 Challenge for releasing high-quality, well-annotated data and for holding such a great challenge to advance the field of image-guided adaptive radiotherapy. We also thank the Center for High Throughput Computing (CHTC) at University of Wisconsin-Madison for providing GPU resources. Dr. Tyler Bradshaw is currently funded by the National Institute Of Biomedical Imaging And Bioengineering of the National Institutes of Health under Award Number R01EB033782.

Disclaimer: The content is solely the responsibility of the authors and does not necessarily represent the official views of the National Institutes of Health.

**Disclosure of Interests.** The authors have no competing interests to declare that are relevant to this article.